\documentclass[prb,twocolumn,notitlepage,longbibliography]{revtex4-2}
\usepackage{amsmath}
\usepackage{amssymb}
\usepackage{bbold} 
\usepackage{natbib}
\usepackage{lipsum}
\usepackage[unicode=true,colorlinks=true,citecolor=blue,urlcolor=blue]{hyperref}
\usepackage{bm}
\usepackage{epsfig} 
\usepackage{tikz} 
\usepackage[normalem]{ulem}

\newcommand {\rme}{{\rm e}}
\begin{document}
\title{Giant PhotoMagnetoDiode Effect}

\author{V.\,L.\,Korenev}
\affiliation{Ioffe Institute, St. Petersburg 194021, Russia}

\author{S.\,A.\,Tarasenko}
\affiliation{Ioffe Institute, St. Petersburg 194021, Russia}

\email{}

\begin{abstract}
We report the observation of a photomagnetodiode effect with a giant rectification ratio of 10 at a magnetic field of 0.1 T.
The effect consists of a diode-like dependence  of the photocurrent on the applied voltage that emerges in an external magnetic field. Such a pronounced nonreciprocity is observed in high-quality VPE-grown GaAs samples at low temperatures at interband photoexcitation.
The photoconductivity is invariant upon reversing the polarity of both electric and magnetic fields. Similar behavior with even higher asymmetry ratio is observed in photoluminescence. The findings are well described by the theory of ambipolar drift of carriers in electric and magnetic fields together with fast surface recombination.
\end{abstract}
\date{\today}

\maketitle
%%%%%%%%%%%%%%%%%%%%%%%%%%%%%%%%%%%%

\section{Introduction}

Nonreciprocity of electron transport in systems with broken time-reversal and space-inversion symmetries has been attracting attention for decades, see review~\cite{Tokura2018}. 
The combined action of electric and magnetic fields on charge carriers has been shown to give rise to a variety of nonreciprocal transport phenomena including 
electrical magnetochiral anisotropy~\cite{Atzori2021,Golub2023,Suarez-Rodriguez2025}, magnetophotogalvanic effect~\cite{Belkov2005,Matsubara2022,Moldavskaya2024}, magnetic quantum ratchet effect~\cite{Falko1989,Tarasenko2011,Drexler2013,Bran2018}, etc. In the case of optical generation of free carriers in semiconductors, the Hall deflection of carriers diffusing from the surface underlies the classical photomagnetoelectric (PME) effect, observed by Kikoin and Noskov as early as 1933 in Cu$_2$O~\cite{Kikoin1934}. In their experiment, a magnetic field $\bm B$ applied in the sample plane generated a photocurrent whose direction reversed upon reversal of the magnetic field. Subsequently, the physics of the PME effect was substantially developed, particularly for anisotropic semiconductors
and $p$-$n$ junctions~\cite{Kikoin1978,Nowak1987}, carriers with momentum alignment
created by linearly polarized light and spin-polarized carriers created by circularly polarized light~\cite{Bakun1984,Alperovich1989,Schmidt2015,Schmidt2017,Dresler2025}, and two-dimensional systems~\cite{Durnev2021}. 
Of special relevance to the present work is the observation of the electric-field‑assisted PME effect in semi-insulating GaAs, where an external in-plane electric field $\bm E$ was superimposed to drive a current with nonlinear I-V characteristics~\cite{Cristoloveanu1984}. 
In early studies of bulk materials, the asymmetry of the I-V characteristics remained modest, requiring the application of a high magnetic field of $12\,$T and a high electric field of $0.6\,$kV/cm to achieve the rectification ratio $R$ of 2~\cite{Cristoloveanu1984}.
Recent studies of nonreciprocal electron transport in two-dimensional systems have demonstrated the asymmetry factor $(R-1)$ of about $10^{-4}$ at polar-magnetic GdTiO$_3$/EuTiO$_3$ interfaces~\cite{Takahara2026} and a light-induced enhancement of the asymmetry factor $(R-1)$ to $\sim 0.1$ at superconducting epitaxial CaZrO$_3$/KTaO$_3$ interfaces~\cite{Zhang2024} at the field of $1~$T.

Here, we report the observation of a 
photomagnetodiode effect (PMDE) with giant magnetic-field-controlled nonreciprocity. While the field dependence of photoconductivity is not new, what is fundamentally different is the regime with pronounced diode‑like I-V characteristics. In high‑quality GaAs samples at interband optical excitation at low temperatures, an extremely high rectification ratio $R$ of 10 is achieved in the magnetic field $B = 0.1\,$T and the electric field $E = 10\,$V/cm. Thus, a relatively small magnetic field, accessible with a standard resistive magnet, transforms a symmetric photoconductivity response into a pronounced diode‑like I-V characteristic. Switching the polarity of the magnetic field reverses the direction of rectification.
Overall, the rectification is controlled by the triple product
$\bm n \cdot (\bm E \times \bm B)$, where
$\bm n$ is the sample normal. The I-V characteristics are determined by the combined action of the electric and magnetic fields and are invariant under simultaneous reversal of the both fields.  
The experimental data are well described both qualitatively and quantitatively by the theory of ambipolar Hall drift of carriers in the crossed electric and magnetic fields together with fast surface recombination. The surface acts as an efficient carrier annihilator, while the magnetic field controls the direction of the Hall drift toward or away from the surface thereby drastically decreasing or increasing the carrier lifetime and density. It is the combination of the high mobility of photocarriers in pure GaAs~\cite{Ruzicka2010}, the ambipolar character of the Hall drift, and fast surface recombination that enables one to achieve a high ratio of the ambipolar Hall-drift length $l_{H,E}$ to the ambipolar diffusion length $l_a$ and, hence, a high rectification ratio at relatively small electric and magnetic fields. The key role of the surface is confirmed by experiments on a GaAs/AlGaAs heterostructure, where the carriers photoexcited in GaAs are separated  from the surface by the AlGaAs barrier. In such a structure, no noticeable asymmetry of the I-V characteristic is observed.

This picture is further supported by the observation of a similar 
$\bm n \cdot (\bm E \times \bm B)$ 
symmetry in photoluminescence (PL). The PL intensity is enhanced or quenched depending on the direction of the Hall drift of photogenerated carriers. Moreover, the PL intensity asymmetry turns out to be even larger than the electrical one, reaching a ratio of about 27. Thus, the electrical and optical responses demonstrate the same physical mechanism of controlling the lifetime and density of photo-injected carriers.

Our results that relatively low magnetic fields can induce giant nonlinearities in the electric and optical properties of semiconductor structures can find applications in optics and optoelectronics.

\section{Experiment}

\begin{figure*}
    \centering
    \includegraphics[width=0.9\linewidth]{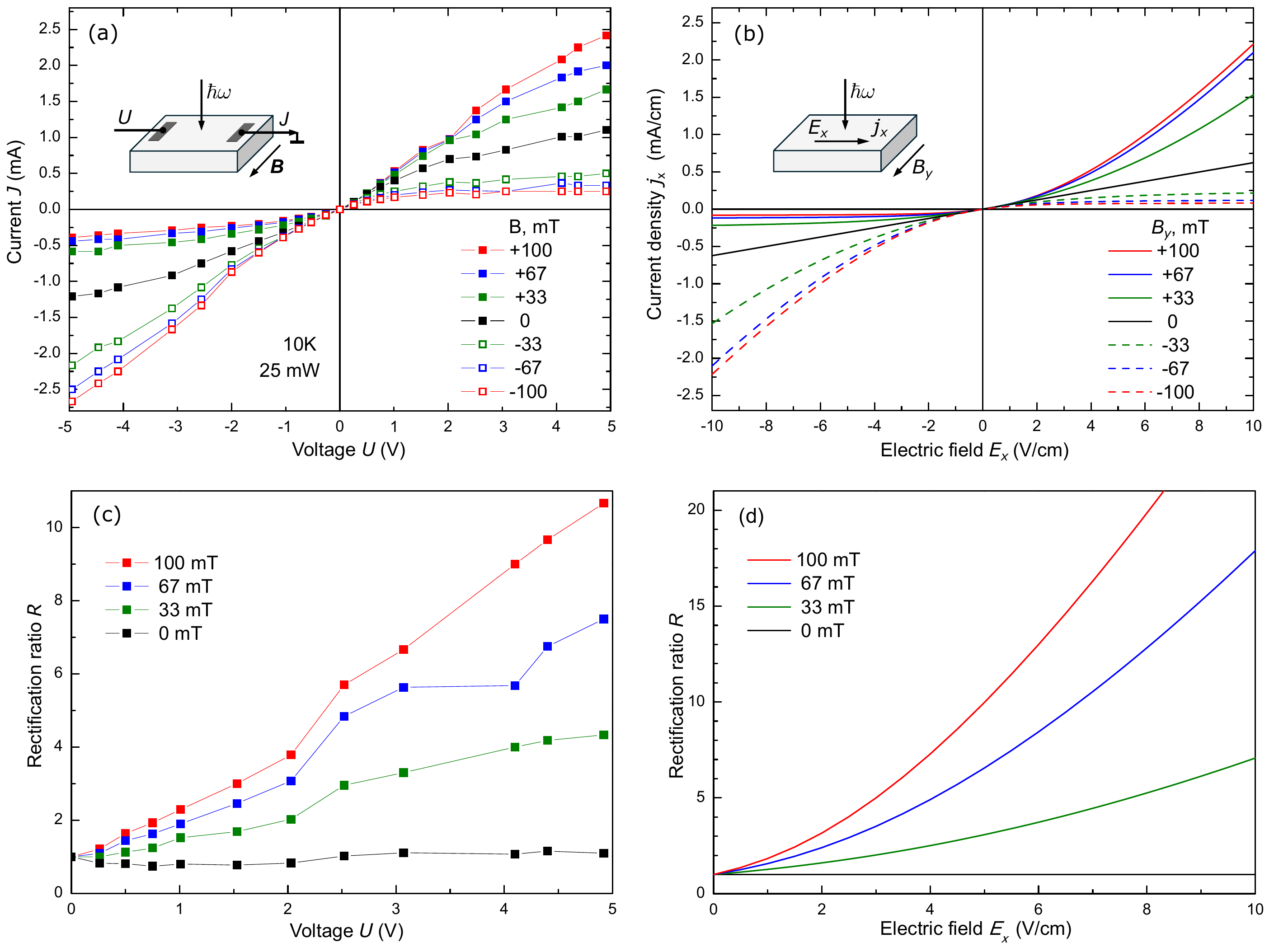}
    \caption{(a) Current-voltage characteristics measured at cw optical excitation at different magnetic fields. Inset shows the experimental geometry. (b) Dependence of current density $j_x$ on the in-plane electric field $E_x$ at different magnetic fields $B_y$ calculated after Eq.~\eqref{J_x_main}. (c) and (d) Rectification ratio $R$ obtained from experimental data and theory, respectively.}
    \label{fig:current} 
\end{figure*}       

The experiments were performed on high-quality bulk GaAs samples. The GaAs layers, grown by vapor-phase epitaxy on (001)-oriented semi-insulating substrates, had a thickness of 30-40 $\mu$m.  The layers were $n$-type, with the difference between the shallow donor and acceptor densities $N_d - N_a  = 6 \times 10^{13}\,$cm$^{-3}$ and the electron mobility $1.4\times10^{5}\,$cm$^2$/(V$\,$s) at $T=77\,$K.
The samples had electrical contacts in the form of two parallel strips, each $2\,$mm long, separated by a distance $d = 5\,$mm. Measurements were carried out at the temperature $T=10\,$K in a closed-cycle cryostat. At these conditions, most resident electrons are bound to shallow donors, and the majority of free carriers in the conduction and valence bands are generated optically. 
The samples were illuminated with a continuous-wave Ti:sapphire laser operating at a wavelength of $750\,$nm, corresponding to the photon energy of $1.65\,$eV, with an optical power of $25\,$mW. The laser beam was focused onto the sample surface using a cylindrical lens to produce an elliptical spot covering the region between the contacts with approximate dimensions of $5\times2$ mm$^2$.

\begin{figure*}
    \centering
    \includegraphics[width=0.9\linewidth]{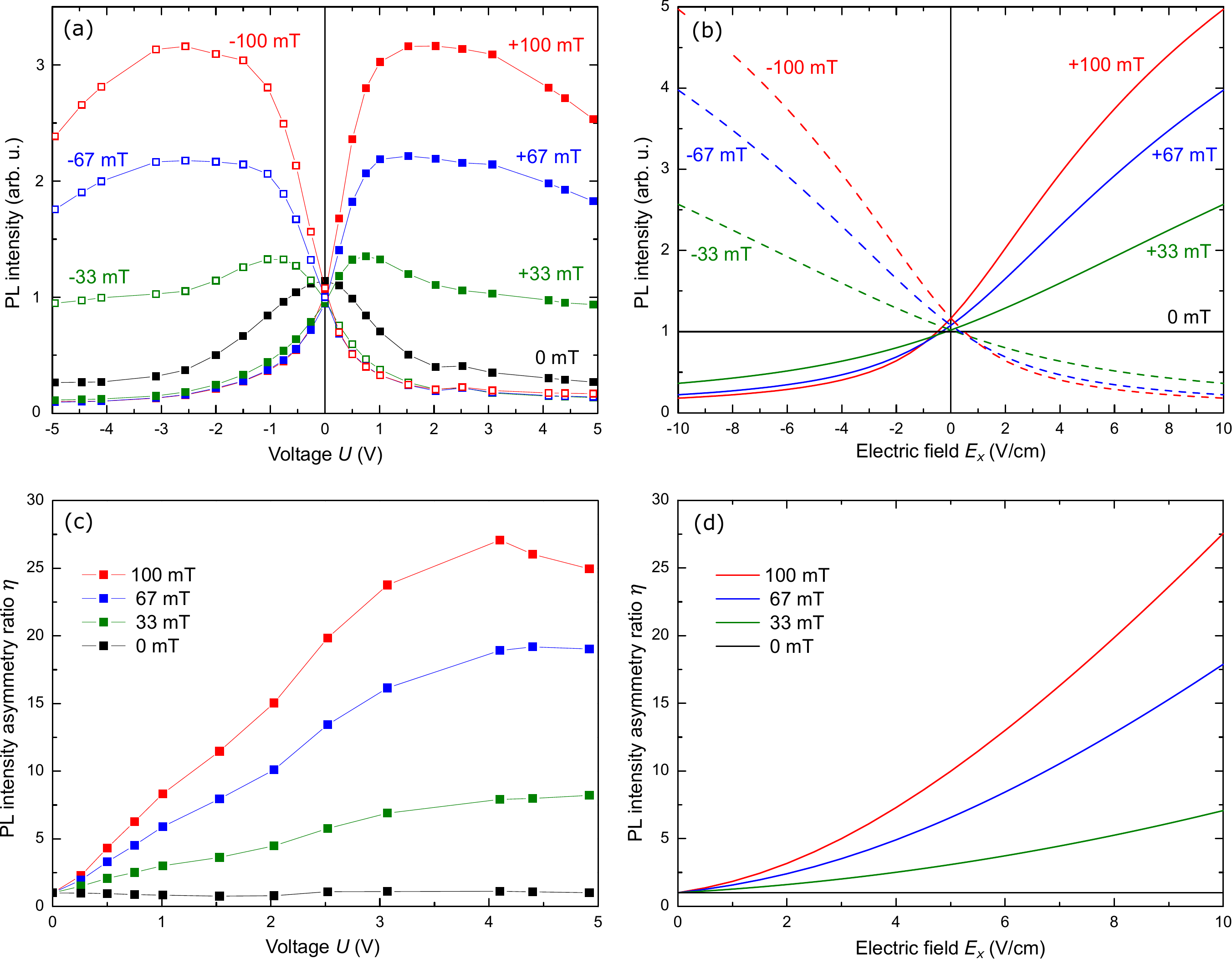}
    \caption{(a) PL intensity as a function the applied voltage $U$ measured at different magnetic fields. (b) Calculated PL intensity as a function the in-plane electric field $E_x$ at different magnetic fields. (c) and (d) PL intensity asymmetry ratio $\eta$ obtained from experimental data and theory, respectively.}
    \label{fig:PL} 
\end{figure*}

Illuminating the sample and applying voltage $U$ between the contacts we detect an electric current $J$. Figure~\ref{fig:current}(a) shows the current–voltage characteristics measured at different values of a magnetic field in the range from $-100\,$mT (open symbols) to $+100\,$mT (solid symbols). The magnetic field $\bm B$ is applied in the surface plane perpendicularly to the current, see inset in Fig.~\ref{fig:current}(a). The voltage $U$ from $-5\,$V to $+5\,$V corresponds to the lateral electric field $E = U/d$ from $-10\,$V/cm to $+10\,$V/cm, respectively.
As the magnetic field $\bm B$ increases, a pronounced diode-like asymmetry develops in the I-V characteristic. Interestingly, the direction of easy current flow is controlled by the magnetic field polarity. Moreover, the I–V curves are invariant under the simultaneous inversion of the voltage $U$ and the field $\bm B$. In other words, the 
photoconductivity exhibits the Hall $\bm E$$\times$$\bm B$ symmetry: it depends on crossed electric $\bm E$ and magnetic $\bm B$ fields but persists upon the inversion of both fields. For collinear $\bm E$ and $\bm B$ fields (not shown), the asymmetry in the I-V curves is absent. 

The asymmetry of the current–voltage characteristics is quantified by the rectification ratio 
\begin{equation}
R = \left| \frac{J(+U,B)}{J(-U,B)} \right|    \,.
\end{equation}
Figure~\ref{fig:current}(c) shows the rectification ratio as a function of the applied voltage and the magnetic field obtained from the experimental data. Strikingly, the asymmetry of the current–voltage characteristics is quite large and the rectification ratio rises from 1 at zero magnetic field up to 10 already in moderate laboratory magnetic fields of $100\,$mT. 

The I–V characteristics are not the only observables demonstrating strong asymmetry in the magnetic field. Similar behavior is observed in a seemingly unrelated experiment on photoluminescence (PL). As is known, the PL of a pure bulk GaAs is dominated by exciton-impurity recombination. Figure~\ref{fig:PL}(a) shows the dependence of the PL intensity on the voltage $U$ for different magnetic fields.  The detection wavelength of $819\,$nm (D$^+$X line) corresponds to the recombination of excitons localized on charged donors. Similarly to the I–V characteristics, the PL intensity as a function of the voltage shows a pronounced  asymmetry which is odd in the magnetic field.  Similar results are obtained for all other major spectral lines, including D$^0$X, D$^0$h, and X. The observed behavior of the PL intensity as a function the $\bm E$ and $\bm B$ fields is universal across the entire spectrum.

Similarly to the rectification ratio for the current, we introduce the PL intensity asymmetry ratio 
\begin{equation}
    \eta = \frac{I_{\rm PL}(+U,B)}{I_{\rm PL}(-U,B)} \,.
\end{equation}
Figure~\ref{fig:PL}(c) shows the asymmetry ratio $\eta$ as a function of the applied voltage and the
magnetic field. Asymmetry in the PL intensity is even more pronounced than that in the current and reaches the ratio of 27.

To conclude the experimental part  of the paper we note that the positive signs of the magnetic field and the voltage in Fig.~\ref{fig:current}(a) and Fig.~\ref{fig:PL}(a), where the current and the PL intensity rise,  correspond to the case where electrons and holes experience drift in the crossed $\bm E$ and $\bm B$ fields from the surface to the sample bulk. Reversing the polarity of either electric or magnetic field causes the carriers to drift toward the surface. This observation indicates the key role of the surface in the photomagnetodiode effect under study. Additionally, we have performed experiment on a MBE-grown GaAs/AlGaAs heterostructure, where the GaAs layer is epitaxially covered by the AlGaAs layer and the photoexcited carriers in GaAs are reliably separated from the sample surface by the AlGaAs barrier. No asymmetry of the current-voltage characteristics is detected in such a sample.

\section{Model and Theory}

\begin{figure}
    \centering
    \includegraphics[width=\linewidth]{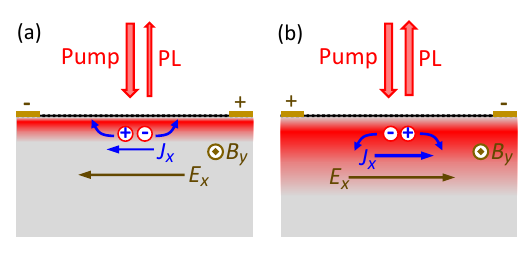}
    \caption{Microscopic mechanism of the photomagnetodiode effect. Drift of photogenerated electron and holes in the in-plane electric $E_x$ and magnetic $B_y$ fields to the surface (panel a) or from the surface (panel b) together with fast surface recombination leads to the depletion or accumulation of carriers. As a result, the photocurrent $J_x$ as a function of the electric field $E_x$ exhibits diode behavior. The photoconductivity is 
    invariant upon switching the polarity of both electric and magnetic fields.}
    \label{fig:model}
\end{figure} 

The microscopic mechanism of the photomagnetodiode effect is sketched in Fig.~\ref{fig:model}. The above-band-gap irradiation generates electrons and holes in the surface layer of the absorption length thickness. Then, the carriers experience drift-diffusion motion in the presence of in-plane electric $E_x$ and magnetic $B_y$ fields. For a given polarity of the fields [Fig.~\ref{fig:model}(a)], the drift in the crossed electric and magnetic fields pushes both types of carriers to the surface. At the surface, they are quickly captured on defect sites, such as dangling bonds, and recombine (primarily non-radiatively). 
The decrease of the free carrier density leads to a decrease of the surface photocurrent $J_x$. The efficient non-radiative surface recombination 
quenches also the PL coming from the bulk radiative recombination.
Switching the polarity of the in-plane electric $E_x$ reverses the situation, Fig.~\ref{fig:model}(b). Now, the drift in the crossed electric and magnetic fields pushes the carriers from the surface, where their lifetime is longer and limited by bulk recombination processes. This leads to the rise of the carrier density, the photocurrent $J_x$, and the PL intensity. Thus, the photoconductivity is sensitive to the electric and magnetic fields but  remains invariant upon switching the polarity 
of both fields.

The theory of the PMDE is presented in Appendix following the drift-diffusion approach of 
Ref.~\cite{Cristoloveanu1984} . Here, we sketch the key equations and discuss the main results. We assume that carriers localized on donors and acceptors do not contribute to transport and consider an ambipolar drift-diffusion model of free carriers. The steady-state distribution of mobile electrons and holes in the depth of the crystal $n(z)$ is found from the ambipolar drift-diffusion equation
\begin{equation}\label{drift-diff-amb}
\mu_a^H E_x \frac{d n}{d z}- D_a \frac{d^2 n}{d z^2} = g(z) - 
\frac{n}{\tau_0} \,,
\end{equation}
where $\mu_a^H$ and $D_a$ are the effective ambipolar Hall mobility 
and diffusion coefficient defined as 
\begin{equation}\label{mua_H}
\mu_a^H = (\theta_e + \theta_h) \frac{\mu_e \mu_h /[(1+ \theta_e^2)(1+ \theta_h^2)]}{\mu_e/(1+\theta_e^2) + \mu_h / (1+ \theta_h^2)} \,,
\end{equation}
\begin{equation}\label{Da}
D_a = \frac{(D_e \mu_h + D_h \mu_e) /[(1+ \theta_e^2)(1+ \theta_h^2)]}{\mu_e/(1+\theta_e^2) + \mu_h / (1+ \theta_h^2)} \,,
\end{equation}
$\mu_{e/h}$ and $D_{e/h}$ are the electron/hole mobilities and diffusion coefficients, respectively, $\theta_{e/h} = \mu_{e/h} B_y /c$ are the Hall angles,
$g(z)$ is the generation rate of electron-hole pairs by the incident light, 
and $\tau_0$ is the lifetime of carriers in the bulk. 

Solution of Eq.~\eqref{drift-diff-amb} for $g(z) = \alpha g_0 \exp(- \alpha z )$, where $\alpha$ is the absorption coefficient, and the boundary condition
$n(0) = 0$, which corresponds to fast surface recombination of carriers, has the form
\begin{equation}\label{n_z_main}
n(z) = \frac{\alpha g_0 \tau_0}{(1-\alpha l_1)(1+\alpha l_2)}
 \left[ \rme^{-\alpha z} - \rme^{- z/l_1}\right] \,,
\end{equation}
where $l_{1,2} = \sqrt{l_a^2 + (l_{H,E}/2)^2} \pm l_{H,E}/2 $, $l_a = \sqrt{D_a \tau_0}$ is the ambipolar diffusion length (affected by the magnetic field) and $l_{H,E} = \mu_a^H \tau_0 E_x$ is the length of ambipolar Hall drift.

Then, the surface density of electrons and holes, which determines the PL intensity, 
$N = \int n(z) dz$ assumes the form
\begin{equation}\label{N_2D}
N  = \frac{g_0 \tau_0}{1+\alpha l_2} \,.
\end{equation}
The surface density of electric current is given by
\begin{equation}\label{J_x_main}
j_x = e\left[ \frac{\mu_e N}{1+\theta_{e}^2} + \frac{\mu_h N}{1+\theta_{h}^2}  + 
\left( \frac{\theta_e \mu_e N}{1+\theta_e^2} - \frac{\theta_h \mu_h N}{1+\theta_h^2} \right) \Theta \right] E_x \,,
\end{equation}
where
\begin{equation}\label{Theta}
\Theta = \frac{\theta_e \mu_e / (1+ \theta_e^2) - \theta_h \mu_h /(1+ \theta_h^2)}{\mu_e / (1+ \theta_e^2) + \mu_h / (1+ \theta_h^2)} \,.
\end{equation}
The last term in Eq.~\eqref{J_x_main} represents the Hall contribution from the transverse electric field $E_z$ emerging near the surface due to different drift of electrons and holes in the applied crossed fields $E_x$ and $B_y$.

Figures~\ref{fig:current}(b), \ref{fig:current}(d), \ref{fig:PL}(b), and~\ref{fig:PL}(d) show the calculated dependences of the current density $j_x$, the rectification ratio $R$, the PL intensity $I_{\rm PL} \propto N$, and the PL intensity asymmetry ratio $\eta$, respectively. All the curves are calculated for the same set of parameters relevant to our GaAs samples and experimental conditions: the absorption length $\alpha = 10^4\,$cm$^{-1}$, the radiation intensity $I = 250\,$mW/cm$^2$ ($25\,$mW per $0.1\,$cm$^2$), the recombination time $\tau_0 = 20\,$ns, the electron and hole mobilities $2\times10^{5}\,$cm$^2$/(V$\,$s) and $\mu_h = \mu_e /10$, respectively, the diffusion coefficients $D_{e/h}$ given by the Einstein relation $D_{e/h} = \mu_{e/h} k_B T /e$, and the carrier temperature $T = 15\,$K. 
The theory reproduces quite well both qualitatively and quantitatively the key experimental observations: the current magnitude and strong asymmetry of the $I-V$ characteristics in the magnetic field, cf. Figs.~\ref{fig:current}(a) and~\ref{fig:current}(b), as well as the behavior of the PL intensity in the electric and magnetic fields, cf. Figs.~\ref{fig:PL}(a) and~\ref{fig:PL}(b). The theory gives higher rectification ratio, cf. Figs.~\ref{fig:current}(c) and~\ref{fig:current}(d), and slightly lower PL intensity asymmetry ratio, cf. Figs.~\ref{fig:PL}(c) and~\ref{fig:PL}(d). Also, it does not reproduce the overall PL intensity decrease at high voltages, Figs.~\ref{fig:PL}(a) and~\ref{fig:PL}(b). We attribute it mainly to the heating of carriers in strong electric fields and the effect of heating on electron transport and recombination processes, which is not included in the theory. 

The theory shows that the dependence of the rectification ratio $R$ on the magnetic field $B_y$ is nonmonotonic. At small magnetic fields, the rectification ratio is given by $R = 1 + \alpha \tau_0 \mu_e \mu_h E_x B_y /[c(1+ \alpha l_0)]$, where 
$l_0 = \sqrt{\tau_0(D_e \mu_h + D_h \mu_e)/(\mu_e + \mu_h)}$ is the zero-field ambipolar diffusion length. At very high magnetic fields, the rectification ratio returns to unity as $R = 1 + \alpha \tau_0 c E_x/B_y$. The highest rectification ratio of the order of $\alpha \tau_0 \sqrt{\mu_e \mu_h} E_x$ is achieved in the intermediate magnetic field $B_m \sim c/\sqrt{\mu_e \mu_h}$, which corresponds to
$\theta_e \theta_h \sim 1$. For the parameters given above, $c/\sqrt{\mu_e \mu_h} \approx 150\,$mT.
The experimentally used field $B = 0.1~$T is close to the theoretically optimal field for the sample parameters.

To summarize, we have demonstrated a regime of magnetic-field-controlled giant nonreciprocity of photocurrent and photoluminescence in GaAs films. Pronounced diode-like I-V characteristics with the rectification ratio of 10 is observed already at a magnetic field of $0.1~$T. The observed photomagnetodiode effect arises from the ambipolar Hall drift of high‑mobility photocarriers in crossed electric and magnetic fields and  fast surface recombination. Pronounced rectification occurs when the ambipolar Hall-drift length becomes comparable to or exceeds the ambipolar diffusion length. 
Thus, even moderate magnetic fields, produced by a standard resistive magnet, can lead to giant nonreciprocity in phototransport when the direction of Hall drift direction is coupled to a spatially localized recombination channel. A similar effect, the edge photomagnetodiode effect, can be expected in two-dimensional structures with fast edge recombination. 

\acknowledgments

We are grateful to Dr. Olga Ken for assistance with experimental work and discussions.

\begin{widetext}

\appendix

\section{Appendix}

We consider a direct-gap semiconductor in the half-space $z > 0$ illuminated by light which creates electron-hole pairs with the generation rate $g(z)$. The semiconductor is
subject to external electric $E_x$ and magnetic $B_y$ fields. The steady-state densities of carries $n_{e/h}(z)$ and densities of fluxes $\bm i_{e/h}(z)$ are found from the equation set
\begin{equation}\label{cont}
\nabla \cdot \bm i_{e/h}  = g(z) - \frac{n_{e/h}}{\tau_{e/h}}  \,,
\end{equation}
\begin{equation}\label{fluxes_B}
\begin{pmatrix} 
\bm i_{e/h,x} \\ \bm i_{e/h,z} 
\end{pmatrix}  
=
\frac{1}{1+\theta_{e/h}^2}
\begin{pmatrix} 
1 & \pm \theta_{e/h} \\  \mp \theta_{e/h} & 1  
\end{pmatrix} 
\begin{pmatrix} 
\mp \mu_{e/h} n_{e/h}  E_x \\ 
- D_{e/h}\, d n_{e/h} dz \mp \mu_{e/h} n_{e/h}  E_z
\end{pmatrix}   \,,
\end{equation}
where $\tau_{e/h}$, $D_{e/h}$, $\mu_{e/h}$, and $\theta_{e/h} = \mu_{e/h} B_y /c$ 
are the electron/hole recombination lifetimes, diffusion coefficients, mobilities, and the Hall angles, respectively.

In the steady-state regime, the charge conservation implies $n_e/ \tau_e = n_h / \tau_h$ and $i_{e,z} = i_{h,z}$. Both conditions are met simultaneously due to formation of immobile charges, such as ionized donors and acceptors, which contribute to the electric field $E_z$. The conditions give
\begin{equation}\label{Ez_B}
E_z = \dfrac{ (\theta_e \mu_e' n_e - \theta_h \mu_h' n_h)E_x + D_h' \, d n_h /dz - D_e' \, d n_e / dz }{ \mu_e' n_e + \mu_h' n_h } 
 = \frac{\theta_e \mu_e' \tau_e - \theta_h \mu_h' \tau_h}{\mu_e' \tau_e + \mu_h' \tau_h} E_x 
+ \frac{D_h' \tau_h - D_e' \tau_e}{\mu_e' \tau_e + \mu_h' \tau_h} \frac{1}{n_e} \frac{d n_e}{d z}  \,,
\end{equation}
where  $\mu_{e/h}' = \mu_{e/h} / (1+ \theta_{e/h}^2)$ and $D_{e/h}' = D_{e/h} / (1+ \theta_{e/h}^2)$. 
Then, equations for the electron/hole distributions assume the drift-diffusion form
\begin{equation}\label{drift-diff}
\mu_{e/h}^H E_x \frac{d n_{e/h}}{d z}- D_{e/h}^a \frac{d^2 n_{e/h}}{d z^2} = g(z) - \frac{n_{e/h}}{\tau_{e/h}} \,,
\end{equation}
where $\mu_{e/h}^{H} =  (\theta_e + \theta_h) \tau_{h/e}  \mu'_e \mu'_h / (\mu'_e \tau_e + \mu'_h \tau_h)$ and
$D_{e/h}^{a} = \tau_{h/e} (D_e' \mu _h' + D_h' \mu_e')/(\mu'_e \tau_e + \mu'_h \tau_h)$.
For the equal recombination times $\tau_e$ and $\tau_h$, Eq.~\eqref{drift-diff} yields Eq.~\eqref{drift-diff-amb} of the main text. 

Solution of Eq.~\eqref{drift-diff} for $g(z) = \alpha g_0 \exp(- \alpha z )$, where $\alpha$ is the absorption coefficient, and the boundary condition
$n_{e/h}(0) = 0$, which corresponds to fast surface recombination of carriers, has the form
\begin{equation}\label{n_z_B}
n_{e/h}(z) = \frac{\alpha g_0 \tau_{e/h}}{(1-\alpha l_1)(1+\alpha l_2)}
 \left[ \rme^{-\alpha z} - \rme^{- z/l_1}\right] \,,
\end{equation}
where 
\begin{equation}\label{l12}
l_{1,2} = \sqrt{D_{e/h}^a \tau_{e/h} + (\mu_{e/h}^H \tau_{e/h} E_x /2)^2}  \pm \mu_{e/h}^H \tau_{e/h} E_x /2 \,.
\end{equation}

It follows that the 2D densities of carriers $N_{e/h} = \int n_{e/h} (z) dz$ and the intensity of PL from the bulk recombination are given by
\begin{equation}\label{N_PL_B}
N_{e/h}  = \frac{g_0 \tau_{e/h}}{1+\alpha l_2}  \,, \;\;  I_{\rm PL} \propto \frac{N_{e}}{\tau_e}  = \frac{N_{h}}{\tau_h} = \frac{g_0}{1+\alpha l_2} \,.
\end{equation}
The effect of electric and magnetic fields on the PL intensity is contained in the dependence of the length $l_2$ on the fields.
For a given magnetic field $B_y > 0$, the length $l_2$ varies from $l_2 \rightarrow \infty$ at $E_x \rightarrow - \infty$ to $l_2 \rightarrow 0$ at 
$E_x \rightarrow +\infty$. Accordingly, the PL intensity drops to zero in the electric field of negative polarity and rises to the maximum possible value in the field of positive polarity.

The density of surface electric current is expressed via the in-plane electron and hole fluxes as follows
\begin{equation}
j_x = e \int [ i_{h,x}(z) - i_{e,x} (z) ] dz \,,
\end{equation}
where the fluxes are given by Eq.~\eqref{fluxes_B}, i.e.,
\begin{equation}
i_{e/h,x} =  \frac{\mp 1}{1+\theta_{e/h}^2} \left[ \mu_{e/h} n_{e/h} E_x + \theta_{e/h} \left( D_{e/h} \frac{d n_{e/h}}{dz} \pm \mu_{e/h} n_{e/h} E_z \right) \right] \,,
\end{equation}
and the field $E_z$ is given by Eq.~\eqref{Ez_B}.  At integrating over the coordinate, 
the terms $\propto \int (d n_{e/h} /dz) dz$ vanish for the boundary conditions $n_{e/h}(0) = 0$.
As a result, the surface electric current assumes the form
\begin{equation}\label{J_x_B}
j_x = e\left[ \frac{\mu_e N_e}{1+\theta_{e}^2} + \frac{\mu_h N_h}{1+\theta_{h}^2}  + 
\left( \frac{\theta_e \mu_e N_e}{1+\theta_e^2} - \frac{\theta_h \mu_h N_h}{1+\theta_h^2} \right)
\frac{\theta_e \mu_e \tau_e / (1+ \theta_e^2) - \theta_h \mu_h \tau_h /(1+ \theta_h^2)}{\mu_e \tau_e / (1+ \theta_e^2) + \mu_h \tau_h / (1+ \theta_h^2)}
\right] E_x .
\end{equation}
It gives Eq.~\eqref{J_x_main} of the main text at $\tau_e = \tau_h$ and $N_e = N_h$.
In our model of fast surface recombination, the surface current is driven by the latter electric field $E_x$ and does not occur at $E_x = 0$ (the Kikoin-Noskov current vanishes).
Equation~\eqref{J_x_B} supplemented with Eq.~\eqref{N_PL_B} for $N_{e/h}$ describes the dependence of the current on the applied electric and magnetic fields.

\end{widetext}


\begin{thebibliography}{99}

\bibitem{Tokura2018} Y. Tokura and N. Nagaosa, Nonreciprocal responses from non-centrosymmetric quantum materials, Nat. Commun. \textbf{9}, 3740 (2018).
https://doi.org/10.1038/s41467-018-05759-4

\bibitem{Atzori2021}
M. Atzori, C. Train, E.A. Hillard, N. Avarvari, G. L. J. A. Rikken, Magneto-chiral anisotropy: From fundamentals to perspectives, Chirality \textbf{33}, 844 (2021).
 https://doi.org/10.1002/chir.23361

\bibitem{Golub2023}
L.E. Golub, E.L. Ivchenko, and B. Spivak, 
Electrical magnetochiral current in tellurium,
Phys. Rev. B \textbf{108}, 245202 (2023).
https://doi.org/10.1103/PhysRevB.108.245202

\bibitem{Suarez-Rodriguez2025}
M. Su\'arez-Rodr\'{\i}guez, B. Mart\'{\i}n-Garc\'{\i}a, F. Calavalle, 
S.S. Tsirkin, I. Souza, F. de Juan, A. Fert, M. Gobbi, L.E. Hueso, and F. Casanova,
Symmetry origin and microscopic mechanism of electrical magnetochiral anisotropy in tellurium, Phys. Rev. B \textbf{111}, 024405 (2025).
https://doi.org/10.1103/PhysRevB.111.024405

\bibitem{Belkov2005}
V.V. Bel'kov, S.D. Ganichev,  E.L. Ivchenko, S.A. Tarasenko,  W. Weber, S. Giglberger,  M. Olteanu, P. Tranitz, S.N. Danilov, P. Schneider, W. Wegscheider, D. Weiss, W. Prettl, Magneto-gyrotropic photogalvanic effects in semiconductor quantum wells, J. Phys.: Condens. Matter \textbf{17}, 3405 (2005).
https://doi.org/10.1088/0953-8984/17/21/032

\bibitem{Matsubara2022}
M. Matsubara, T. Kobayashi, H. Watanabe, Y. Yanase, S. Iwata, and T. Kato,
Polarization-controlled tunable directional spin-driven photocurrents in a magnetic metamaterial with threefold rotational symmetry, Nat. Commun. \textbf{13}, 6708 (2022). 
https://doi.org/10.1038/s41467-022-34374-7

\bibitem{Moldavskaya2024} 
M.D. Moldavskaya, L.E. Golub, V.V. Bel'kov, S.N. Danilov, D.A. Kozlov, J. Wunderlich, D. Weiss, N.N. Mikhailov, D.A. Dvoretsky, S.S. Krishtopenko, B. Benhamou-Bui, F. Teppe,
and S.D. Ganichev, Magnetophotogalvanic effects driven by terahertz radiation in CdHgTe crystals with Kane fermions, 
Phys. Rev. B \textbf{110}, 205204 (2024).
https://doi.org/10.1103/PhysRevB.110.205204

\bibitem{Falko1989}
V.I. Fal'ko, Rectifying properties of 2D inversion layers in a
parallel magnetic field, Sov. Phys. Solid State \textbf{31}, 561 (1989).

\bibitem{Tarasenko2011}
S.A. Tarasenko, Direct current driven by ac electric field in
quantum wells, Phys. Rev. B \textbf{83}, 035313 (2011).
https://doi.org/10.1103/PhysRevB.83.035313

\bibitem{Drexler2013}
C. Drexler, S.A. Tarasenko, P. Olbrich, J. Karch, M. Hirmer, F. M\"{u}ller, M. Gmitra, J. Fabian, R. Yakimova, S. Lara-Avila, S. Kubatkin, M. Wang, R. Vajtai, P.M. Ajayan, J. Kono, 
and S.D. Ganichev, Magnetic quantum ratchet effect in graphene, Nat. Nanotechnol. \textbf{8}, 104 (2013).
https://doi.org/10.1038/nnano.2012.231

\bibitem{Bran2018}
C. Bran, E. Berganza, J.A. Fernandez-Roldan, E.M. Palmero, J. Meier, E. Calle,  M. Jaafar, M. Foerster, L. Aballe, A.F. Rodriguez, R.P. del Real, A. Asenjo, O. Chubykalo-Fesenko, and M. Vazquez, Magnetization ratchet in cylindrical nanowires,
ACS Nano \textbf{12}, 5932 (2018).
https://doi.org/10.1021/acsnano.8b02153

\bibitem{Kikoin1934}
I.K. Kikoin and M.M. Noskov, A new photoelectric effect in
cuprous oxide, Phys. Z. Sowjetunion \textbf{5}, 586 (1934).

\bibitem{Kikoin1978}
I.K. Kikoin and S.D. Lazarev, 
Photoelectromagnetic effect, Sov. Phys. Usp. \textbf{21}, 297 (1978).
http://dx.doi.org/10.1070/pu1978v021n04abeh005538

\bibitem{Nowak1987}
M. Nowak, 
Photoelectromagnetic effect in semiconductors and its applications,
Prog. Quantum Electron. \textbf{11}, 205 (1987).
https://doi.org/10.1016/0079-6727(87)90001-2

\bibitem{Bakun1984} 
A.A. Bakun, B.P. Zakharchenya, A.A. Rogachev, M.N. Tkachuk, and 
V.G. Fleisher, Observation of a surface
photocurrent caused by optical orientation of electrons in a
semiconductor, JETP Lett. \textbf{40}, 1293 (1984).
http://jetpletters.ru/ps/1262/article\_19087.shtml

\bibitem{Alperovich1989}
V.L. Alperovich, A.O. Minaev, and A.S. Terekhov, Ballistic electron
transport through epitaxial GaAs films in a magnetically induced surface
photocurrent,
JETP Lett. \textbf{49}, 702 (1989).
http://jetpletters.ru/ps/1122/article\_16999.shtml

\bibitem{Schmidt2015}
C.B. Schmidt, S. Priyadarshi, S.A. Tarasenko, M. Bieler, 
Ultrafast magneto-photocurrents in GaAs: Separation of surface and bulk contributions, 
Appl. Phys. Lett. \textbf{106}, 142108 (2015).
https://doi.org/10.1063/1.4917568

\bibitem{Schmidt2017} C.B. Schmidt, S. Priyadarshi, and M. Bieler, 
Sub-picosecond temporal resolution of anomalous Hall currents in GaAs, 
Sci. Rep. \textbf{7}, 11241 (2017).
https://doi.org/10.1038/s41598-017-11603-4

\bibitem{Dresler2025}
C. Dresler, S. Priyadarshi, and M. Bieler,
Magnetic- and electric-field-induced anomalous Hall currents from optical excitation of Landau transitions in bulk GaAs, 
Phys. Rev. B \textbf{111}, 155201 (2025).
https://doi.org/10.1103/PhysRevB.111.155201

\bibitem{Durnev2021}
M.V. Durnev and S.A. Tarasenko, Edge photogalvanic effect caused by optical alignment of carrier momenta in two-dimensional Dirac materials, Phys. Rev. B \textbf{103}, 165411 (2021).
https://doi.org/10.1103/PhysRevB.103.165411

\bibitem{Cristoloveanu1984} S. Cristoloveanu and K.N. Kang, The field-assisted photomagnetoelectric effect: theory and experiment in semi-insulating GaAs, 
J. Phys. C: Solid State Phys. \textbf {17}, 699 (1984).
https://doi.org/10.1088/0022-3719/17/4/012

\bibitem{Takahara2026}
N. Takahara, K.S. Takahashi, N. Nagaosa, Y. Tokura, and M. Kawasaki,
Gigantic Nonreciprocal Conduction at a Polar-Magnetic Interface of GdTiO$_3$/EuTiO$_3$,
Phys. Rev. Lett. \textbf{136}, 206304 (2026).
https://doi.org/10.1103/lfcm-jyb6

\bibitem{Zhang2024}
X. Zhang, T. Zhu, Sh. Zhang, Zh. Chen, A. Song, Ch. Zhang, R. Gao, W. Niu, Y. Chen, F. Fei, Y. Tai, G. Li, B. Ge, W. Lou, J. Shen,
H. Zhang, K. Chang, F. Song, R. Zhang, and X. Wang,
Light-induced giant enhancement of nonreciprocal transport 
at KTaO$_3$-based interfaces,
Nat. Commun. \textbf{15}, 2992 (2024).
https://doi.org/10.1038/s41467-024-47231-6

\bibitem{Ruzicka2010}
B.A. Ruzicka, L.K. Werake, H. Samassekou, and H. Zhao,
Ambipolar diffusion of photoexcited carriers in bulk GaAs, 
Appl. Phys. Lett. \textbf{97}, 262119 (2010).
https://doi.org/10.1063/1.3533664

\end{thebibliography}
\end{document}